# Kinetics of Particles in Relativistic Collisionless Shocks

Mikhail V. Medvedev

*Department of Physics and Astronomy, University of Kansas, Lawrence, Kansas*

Luis O. Silva, Ricardo A. Fonseca

*GoLP/Instituto Superior Tecnico, Lisbon, Portugal*

J. W. Tonge, Warren B. Mori

*Department of Physics and Astronomy, University of California at Los Angeles, Los Angeles, California*

Charged plasma particles form an anisotropic counter-streaming distribution at the front of a shock. In the near- and ultra-relativistic regimes, Weibel (or two-stream) instability produces near-equipartition, chaotic, small-scale magnetic fields. The fields introduce effective collisions and "thermalize" the plasma particles via pitch-angle scattering. The properties of jitter radiation emitted by accelerated electrons from these small-scale fields are markedly different from synchrotron spectra. Here we theoretical summary and the results of recent numerical 3D PIC plasma kinetic simulations. The relation of the obtained results to the theory of gamma-ray bursts is outlined.

## 1. INTRODUCTION

It is well known that hydrodynamic shocks may occur in neutral media, in which collisions between gas particles mediate the pressure. However, the interplanetary space gas is very dilute and highly ionized. Hence Coulomb collisions are very rare and electric and magnetic fields, instead, mediate particle-particle interactions. Explosions in such rarefied plasma also drive shock waves, but these shocks are very different from their hydrodynamic counterparts and are referred to as "collisionless". The structure and properties of collisionless shocks vary dramatically and depend on many parameters, e.g., the value and orientation of magnetic fields in a plasma, the temperatures and masses of different plasma species, the shock speed compared to the speed of light, and others. Recently we made a great progress in theoretical understanding of ultra-relativistic collisionless shock moving in unmagnetized plasma and showed that non-linear Weibel (two-stream) instability plays a major role. A novel plasma radiation mechanism, which we discovered and called the ``jitter'' mechanism, is an unambiguous spectral benchmark of strong Weibel turbulence present in the shock. First three-dimensional particle-in-cell plasma kinetic simulations, performed with the OSIRIS code by the USA--Portugal collaboration, fully confirm our predictions. Moreover, this model receives an independent and very strong observational support from astrophysics, namely spectral characteristics of gamma-ray bursts.

## 2. RELATIVISTIC SHOCK THEORY

In general, relativistic shocks, as well as other shocks with the Mach number greater than three, must be highly turbulent. The source and the mechanism of the turbulence is thought to be kinetic in order to prevent multi-stream motion of plasma particles. It has been shown that the relativistic Weibel instability operates on the shock front [*Medvedev* and *Loeb*, 1999]. This instability is driven by the anisotropy of the particle distribution function (PDF) associated with a large number of particles reflected from the shock potential.

The instability under consideration was first predicted by [*Weibel*, 1959] for a non-relativistic plasma with an anisotropic distribution function. The simple physical interpretation provided later by [*Fried*, 1959] treated the PDF anisotropy more generally as a two-stream configuration of cold plasma. Below we give a brief, qualitative description of this two-stream magnetic instability.

Let us consider, for simplicity, the dynamics of the electrons only, and assume that the protons are at rest and provide global charge neutrality. The electrons are assumed to move along the *x*-axis (as illustrated in Figure 1) with velocities $\mathbf{v}=+\mathbf{x}v_x$ and $\mathbf{v}=-\mathbf{x}v_x$, and equal particle fluxes in opposite directions along the *x*-axis (so that the net current is zero). Next, we add an infinitesimal magnetic field fluctuation, $\mathbf{B}=\mathbf{z}B_z\cos(ky)$. The Lorentz force, $-e(\mathbf{v} \times \mathbf{B})/c$, deflects the electron trajectories. As a result, the electrons moving to the right and those moving to the left will concentrate in spatially separated current filaments. The magnetic field of these filaments appears to increase the initial magnetic field fluctuation. The growth rate is $\Gamma=\omega_p(v_y/c)$. Similar considerations imply that perpendicular electron motions along *y*-axis, result in oppositely directed currents, which suppress the instability. The particle motions along **z** are insignificant as they are unaffected by the magnetic field. Thus, the instability is indeed driven by the PDF anisotropy and should quench for the isotropic case.

The Lorentz force deflection of particle orbits increases as the magnetic field perturbation grows in amplitude. The amplified magnetic field is random in the plane perpendicular to the particle motion, since it is generated from a random seed field. Thus, the Lorentz deflections result in a pitch angle scattering, which makes the PDF isotropic. If one starts from a strong anisotropy, so that the thermal spread is much smaller than the particle bulk velocity, the particles will eventually isotropize and the thermal energy associated with their random motions will be equal to their initial directed kinetic energy. This final state will bring the instability to saturation.

Below we note the following points about the nature of the instability:

(1) The free energy source is the anisotropy of the PDF at and near the shock associated with a two-stream motion of plasma particles: the inflowing

(in the shock frame) particles of the external medium and the group of outgoing particles reflected from the shock front.

(2) Despite its intrinsically kinetic nature, the instability is non-resonant, i.e., it is impossible to single out a group of particles that is responsible for the instability. Since the bulk of the plasma participates in the process, the energy transferred to the magnetic field could be comparable to the total kinetic energy of the plasma.

(3) The characteristic growth time (inverse growth rate) for each species (electron, proton, positron, if any) is $\sim\omega_p^{-1}$ [$\omega_p=(4\pi e^2 n/m)^{1/2}$ is the plasma frequency] in the shock frame, which is very short compared to a shock evolution time scale.

(4) The instability is *aperiodic*, i.e., Re $\omega$=0. Thus, it can be saturated by nonlinear effects only, and not by kinetic effects such as collisionless damping or resonance broadening. Hence, the magnetic field strength may reach a large amplitude.

(5) The instability generates randomly oriented magnetic fields on a spatial scale of order the plasma skin-depth $\sim c/\omega_p$. These fields predominantly lie in the plane of the shock front.

(6) The instability introduces an effective scattering process into the otherwise collisionless system. This validates the use of the MHD approximation in the study of the large-scale dynamics of collisionless shocks. It makes the PDF isotropic via pitch-angle scattering and, thus, effectively heats the electrons and protons.

## 1. NUMERICAL SIMULATIONS OF THE WEIBEL INSTABILITY

The dynamics of the Weibel instability has recently been simulated by several research groups using 3D plasma kinetic codes. We examined the instability, which occurs in a collision of two inter-penetrating unmagnetized plasma blobs with zero net charge [*Silva, et al.,* 2003]. This is the simplest model for the formation region of a shock front, as well as a classic scenario unstable to electromagnetic and/or electrostatic plasma instabilities. To probe the full nonlinear dynamics and the saturated state of this system it is necessary to employ kinetic numerical simulations.

The fully electromagnetic relativistic PIC code OSIRIS [*Fonseca, et al.,* 2002] was used to perform the first three-dimensional kinetic simulations of the collision of two plasma shells, and to observe the three-dimensional features of the electromagnetic filamentation instability, or Weibel instability.

The simulations were performed on a 256x256x100 grid [the box size is 25.6x25.6x10.0 ($c/\omega_{pe}$)$^3$] with the total of 105 million particles for 2900 time steps [corresponding to

150.0 $\omega_{pe}^{-1}$], with periodic boundary conditions. In all runs, energy is conserved down to 0.025%. In our simulations, at $t=0$ there are two groups of particles moving along the vertical (*x3*) axis in opposite directions in the center of mass frame and occupying the entire simulation volume. The particles in both groups have a small thermal spread. The system has no net charge and no net current, and initially the electric and magnetic fields are set to zero. We performed both sub-relativistic ($\gamma_0$~1.17, $\gamma_0$ is the initial Lorentz factor) and ultra-relativistic ($\gamma_0$~10.05) simulations. Note that in the dimensionless units used, the results of the simulations are equally applicable to the collision of electron-proton plasmas as well.

The 3D structure of current filaments is shown in Figure 2. The temporal evolution of the total energy in the produced magnetic and electric fields is shown in Figure 3. During the linear stage of the instability there is rapid generation of a strong magnetic field, which predominantly lies in the plane of the shock (*x1x2*--plane), i.e., perpendicular to the direction of motion of the plasma shells. The magnetic field energy density reaches ~5...20\% for $\gamma_0$~1.17...10.05. In all cases, the produced electric field is significantly weaker than the magnetic field. The linear growth rate agrees well with the theoretical estimates for the Weibel instability.

After a short linear stage, the instability enters the nonlinear regime (at $t$~10$\omega_{pe}^{-1}$) in which current filaments begin to interact with each other, forcing like currents to approach each other and merge. During this phase, initially randomly oriented filaments cross each other to form a more organized, large-scale quasi-regular patterm, hence much current and *B*-field is annihilated. At later times ($t$~30$\omega_{pe}^{-1}$ and later), the filament coalescence continues, as indicated by the increase of the correlation scale, ~$k^{-1}$, of the *B*-field in Figure 4. However, the spatial distribution of currents is now quite regular, so that filaments with opposite polarity no longer cross each other but simply interchange, staying always far away. The total magnetic field energy is ~0.25% and does not change any more. Note that the residual magnetic field is highly inhomogeneous, seen as a collection of magnetic field filaments or "bubbles". The amplitude of the field in the bubbles is close to equipartition. Therefore, the overall decrease of the *B*-field energy is mostly associated with decreasing *filling factor* of the field. Note also that the magnetic domains separate current filaments of opposite polarity.

The topological evolution of the magnetic field is accompanied by heating and non-thermal particle acceleration, as illustrated in Figure 5. The particle energization is due to the pitch angle scattering in the produced *B*-field. The generation of non-thermal fast particles is more pronounced in the highest magnetic field scenarios. The presence of such high-energy particles is fundamental to provide the mildly relativistic particles to be injected in the accelerating structure of a shock that forms in the collision region.

## 4. JITTER RADIATION FROM SHOCKS

There is lore that radiation observed from astrophysical shocks is the synchrotron radiation emitted by accelerated electrons in nearly homogeneous (at least on a Larmor scale) magnetic fields. We have shown that, instead, the magnetic field in Weibel shocks is randomly tangled and its correlation length is less then the Larmor radius of an emitting electron. Hence, such an electron experiences random deflections as it moves through the field and its trajectory is, in general, stochastic. The synchrotron theory fails here.

In collisionless shocks dominated by Weibel turbulence, the particle moves almost straight, along the line of sight almost straight and experiences high-frequency jittering in the perpendicular direction due to the random Lorentz force. We therefore refer the emerging radiation to as ``*jitter*'' radiation. Its spectrum is determined by random accelerations of the particle; it will be peaked at some frequency $\omega_j$, which is estimated as follows. In the rest frame of the electron, the magnetic field inhomogeneity with wavenumber $k_B$ is transformed into a transverse pulse of electromagnetic radiation with frequency $k_B c$. This radiation is then Compton scattered by the electron to produce observed radiation with frequency $\omega_j \sim \gamma^2 k_B c$ in the lab frame. It has been shown [*Medvedev*, 2000] that this frequency, referred to as the jitter frequency, is higher than the synchrotron frequency in the uniform magnetic field of the same strength. The calculated spectrum of jitter radiation is shown in Figure 6. The crucial difference of its spectrum from synchrotron is (i) harder spectrum: the photon spectra flux goes at low energies as $F \sim \nu^\alpha$ with $\alpha$=0, in contrast to synchrotron $\alpha$=–2/3, and (ii) much sharper spectral breat at the jitter frequency. These properties can naturally explain the violation of the "synchrotron line of death" in time-resolved GRB spectra and the nature of the broken power-law spectra of GRBs.

___________

Centro de Fisica de Plasmas, Instituto Superior Tecnico, Av. Rovisco Pais, 1049-001 Lisboa, Portugal

Department of Physics and Astronomy, University of Kansas, 1251 Wescoe Hall Drive, #1082, Lawrence, Kansas, 66045-7582, USA

Department of Physics and Astronomy, University of California at Los Angeles, 405 Hilgard Avenue, Box 951361, Los Angeles, California, 90095-1361, USA

FIGURE CAPTIONS

**Figure 1.** A diagram representing the mechanism of the Weibel instability. Counter-propagating charges are deflected by infinitesimally small magnetic fields to form current filaments. The magnetic fields of the filaments add up to (amplify) the initial magnetic field and create a positive feedback loop.

**Figure 2.** Three-dimensional structure of current filaments obtained from plasma kinetic 3D PIC simulations, at two different times ($t$=10 and $t$=50 electron plasma times; $\gamma\beta$=0.6).

**Figure 3.** Temporal evolution of the magnetic and electric fields normalized to the total initial kinetic energy of plasma streams for sub- and ultra-relativistic shock conditions.

**Figure 4.** Temporal evolution of the magnetic field spectral density distribution, normalized to the peak spectral density for the sub-relativistic shock.

**Figure 5.** The evolution of the particle distribution function.

**Figure 6.** Spectra of jitter radiation; the synchrotron spectrum (dashed curve) is shown for comparison.

PARTICLE KINETICS IN RELATIVISTIC SHOCKS

MEDVEDEV, ET AL.

K:\AU_PACKS\GMNEW\TEMPLATE.WPD

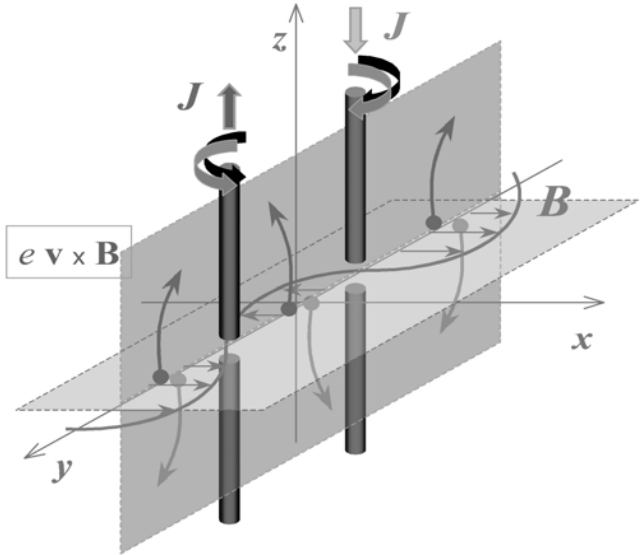

fig 1

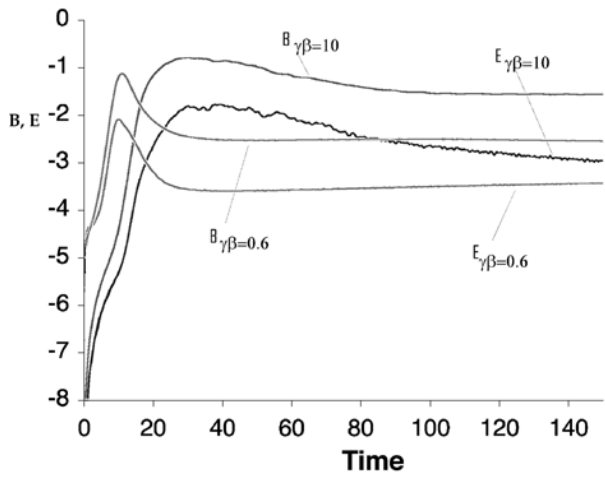

fig 3

$\gamma - 1$

Time Evolution

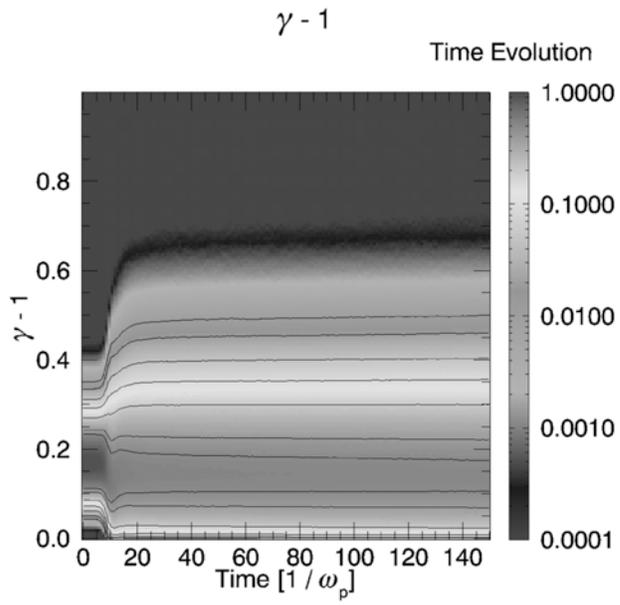

fig 4

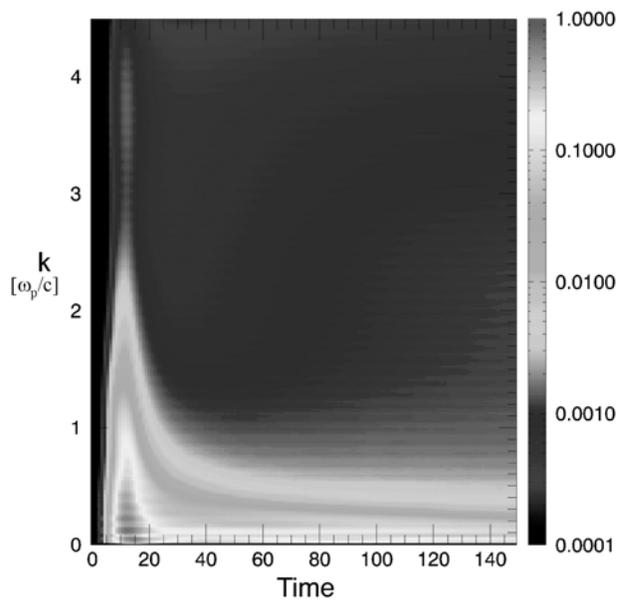

fig 5

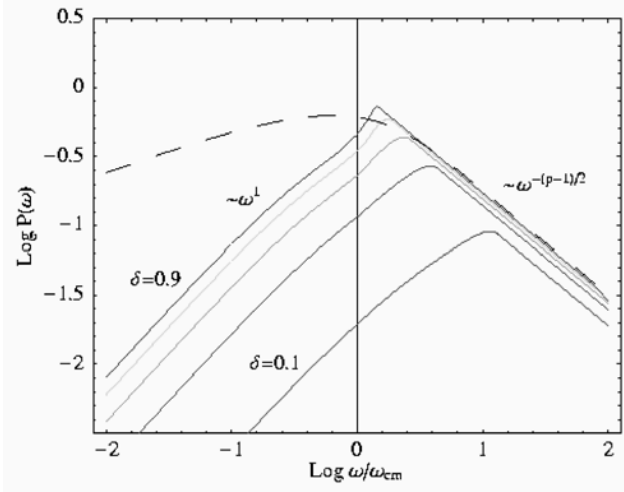

fig 6

Mass Density

Time = 10.40 [ 1 / $\omega_s$ ]

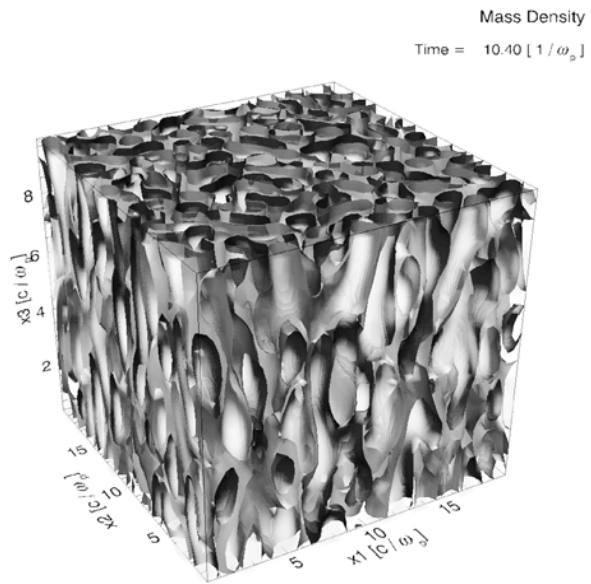

Mass Density

Time = 50.96 [ 1 / $\omega_s$ ]

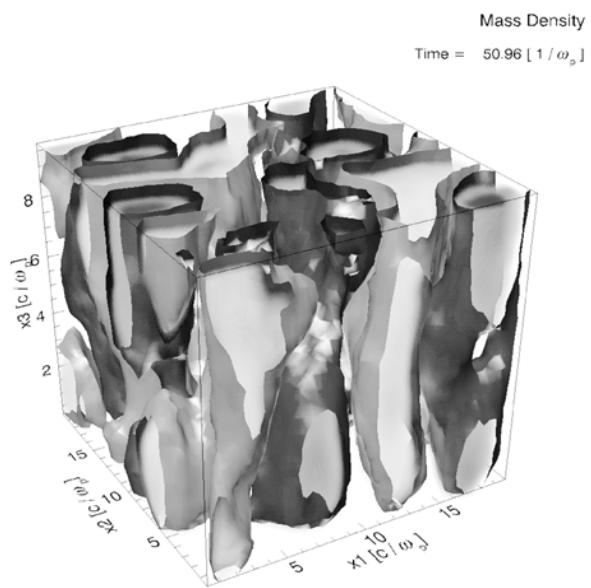

fig 2